\documentclass[12pt]{article}
\usepackage{graphicx,xcolor}
\usepackage[colorlinks,linkcolor=blue,citecolor=blue,urlcolor=blue,bookmarks,bookmarksnumbered]{hyperref}
\usepackage{amsmath,amssymb,amsfonts,mathrsfs}
\usepackage[paper=letterpaper, margin=1.0in, top=1.5in, bottom=0.33in]{geometry}
\usepackage{cite}
\RequirePackage{tikz}
 \usetikzlibrary{arrows}
 
\usepackage{enumitem}

\newcommand{\be}{\begin{equation}}
\newcommand{\ee}{\end{equation}}
\newcommand{\bea}{\begin{eqnarray}}
\newcommand{\eea}{\end{eqnarray}}
\newcommand{\ba}{\begin{array}}
\newcommand{\ea}{\end{array}}
\newcommand{\ben}{\begin{enumerate}}
\newcommand{\een}{\end{enumerate}}
\newcommand{\bi}{\begin{itemize}}
\newcommand{\ei}{\end{itemize}}
\newcommand{\bc}{\begin{center}}
\newcommand{\ec}{\end{center}}
\newcommand{\bfig}{\begin{figure}}
\newcommand{\efig}{\end{figure}}
\newcommand{\bq}{\begin{quotation}}
\newcommand{\eq}{\end{quotation}}
\newcommand{\bt}{\begin{table}}
\newcommand{\et}{\end{table}}
\newcommand{\btab}{\begin{tabular}}
\newcommand{\etab}{\end{tabular}}
\newcommand{\bs}{\begin{slide}}
\newcommand{\es}{\end{slide}}

\newcommand{\pa}{\partial}

\newcommand{\X}{\mathbb{X}}
\newcommand{\cX}{\mathcal{X}}

\newcommand{\bra}{\langle}
\newcommand{\ket}{\rangle}

\newcommand{\vev}[1]{\langle #1 \rangle}

\newdimen\lft\lft=0pt

\newcommand{\beq}{\begin{eqnarray}}
\newcommand{\eeq}{\end{eqnarray}}
\newcommand{\beqn}{\begin{eqnarray}}
\newcommand{\eeqn}{\end{eqnarray}}

\def\pa{\partial}
\newcommand{\rd}{\mathrm{d}}
\newcommand{\rD}{\mathrm{D}}
\newcommand{\ls}{\ell_s}

\let\SS=\S 
\def\S{\mathbb{S}}

\def\bra{\langle}
\def\ket{\rangle}

\let\ba=\overline

\let\d=\delta

\def\define{\buildrel{\rm def}\over=}

\let\j=\psi

\let\L=\Lambda

\let\t=\tau

\def\vev#1{\left\langle#1\right\rangle}

\def\RR{\relax\leavevmode
       \ifmmode\mathchoice
       {\hbox{\sf R\kern-.4em R}}
       {\hbox{\sf R\kern-.4em R}}
       {\lower.9pt\hbox{\scriptsize\sf R\kern-.36em R}}
       {\lower1.2pt\hbox{\tiny\sf R\kern-.36em R}}
       \else{\sf R\kern-.4em R}\fi}

\def\resetby#1#2{\@addtoreset{#2}{#1}}
\def\seceq{\@addtoreset{equation}{section}
              \def\theequation{\thesection.\arabic{equation}}}

\def\Label#1{\label{#1}%
                \smash{\hbox to0pt{\raise1ex\hbox{\tiny[#1]}\hss}}}
\def\noLabels{\let\Label=\label}

\DeclareMathOperator{\Tr}{\textrm{Tr}}

\let\sss=\scriptscriptstyle

 \allowdisplaybreaks
 \numberwithin{equation}{section}
 
 \parskip\medskipamount
\begin{document}
\setcounter{page}{-1}
\thispagestyle{empty}

\begin{center}

\vskip 10mm
\begin{center}\Large \bf
    String Theory Bounds on the Cosmological Constant,
    the Higgs Mass, and the Quark and Lepton Masses
\end{center}
\vskip 10mm

\renewcommand{\thefootnote}{\fnsymbol{footnote}}

\centerline{{\bf
Per Berglund${}^{1}$\footnote{\tt per.berglund@unh.edu},
Tristan H\"{u}bsch${}^{2}$\footnote{\tt thubsch@howard.edu}
and
Djordje Minic${}^{3}$\footnote{\tt dminic@vt.edu}
}}

{\it
${}^1$Department of Physics and Astronomy, University of New Hampshire, Durham, NH 03824, U.S.A. \\
${}^2$Department of Physics and Astronomy, Howard University, Washington, D.C.  20059, U.S.A. \\
${}^3$Department  of Physics, Virginia Tech, Blacksburg, VA 24061, U.S.A. \\
${}$ \\
}

\end{center}

\vskip 5mm

\begin{abstract}
We elaborate on the new understanding of the cosmological constant and the gauge hierarchy problems in the context of string theory
in its metastring formulation, based on the concepts of modular spacetime and Born geometry.
The interplay of phase space (and Born geometry), the Bekenstein bound, the mixing between ultraviolet (UV) and infrared (IR) physics and modular invariance in string theory is emphasized.
This new viewpoint is fundamentally rooted in quantum contextuality and not in 
statistical observer bias (anthropic principle).
We also discuss the extension of this point of view to the problem of masses of quarks and leptons and their respective mixing matrices.
\end{abstract}

\vfill
{\em In memory of Steve Weinberg}
\vfill
\clearpage

\begingroup
\thispagestyle{empty}
\baselineskip=10pt 
\setcounter{tocdepth}{3}
  \begin{center}
    \begin{minipage}{140mm}
      \tableofcontents
    \end{minipage}
  \end{center}
\endgroup

\setcounter{footnote}{0}
\renewcommand{\thefootnote}{\arabic{footnote}}
\vfill
\clearpage

\section{Introduction} 
\label{s:Intro}
As emphasized by Steven Weinberg, physics indeed strives on crisis\cite{Weinberg:1988cp},
and the by now proverbial $10^{120}$-size ``cosmological constant problem'' of Quantum Field Theory (QFT) has been (together with the gauge hierarchy problem) such a vexing conundrum for so long, that it prompted (then) novel and (more recently) very influential ideas, such as the anthropic reasoning\cite{Weinberg:1987dv}
and the string landscape\cite{Polchinski:2006gy}.
Another vexing issue in fundamental physics is the intricate pattern of fermion masses in the Standard Model\cite{Weinberg:2020zba}.
In this contribution to the special issue devoted to
Steve Weinberg we would like to elaborate on 
and extend the discussion 
of a new approach that offers a {\em unified\/} understanding to 
the cosmological constant (``cc'')
and the gauge hierarchy (Higgs mass) problems
following our recent work
\cite{Freidel:2022ryr, Berglund:2022qsb}. Then we would like to extend this approach to some intriguing observations regarding the problem of fermion masses\cite{Minic:2023oty}\footnote{This article is based on the recent talk given at the CERN conference on the exotic approaches to the hierarchy problems, as well as the talk given at BASIC 2023\cite{Minic:2023oty} and the lecture on the ``Challenges of Quantum Gravity'' sponsored by the International Society for Quantum Gravity.}.

The central issue of our article is the ultraviolet (UV)/infrared (IR) mixing (a feature that goes beyond effective field theory (EFT)) and quantum contextuality 
(as opposed to anthropic reasoning) in the vacuum energy problem. We also stress
the role of phase-space and modular polarization (endowed with Born geometry\cite{Freidel:2013zga} --- a kind of generalized 
mirror symmetry\cite{Berglund:2021hbo}) and the Bekenstein\cite{Bekenstein:1980jp}, or holographic\cite{tHooft:1993dmi,Susskind:1994vu,Fischler:1998st} bound, 
combined with stringy modular invariance
\cite{Polchinski:1998rq, Polchinski:1998rr}
and a stringy formula for the Higgs mass\cite{Abel:2021tyt}. Furthermore, 
we emphasize the overall 
relevance of these concepts for addressing the problem 
of masses and mixing matrices of quarks and leptons
in the context of string theory/quantum gravity 
connecting to the recent work on Bjorken \cite{Bjorken:2013aa,rBJ-MM}. In this last instance we make some tantalizing observations that call for more definitive future calculations.
Finally, we emphasize that 
these issues 
are all realized in the framework of string theory, a consistent theory of quantum gravity and Standard Model-like matter, justifying the title of the present paper.

The paper is organized as follows:
Section~\ref{s:CC} discusses various facets of the cosmological constant (cc) problem,
reviewing it first (\SS\,\ref{s:CCP}) in point-particle QFT, 
and then also in string theory.
\SS\,\ref{s:CC-fix} then presents the recent resolution of this problem \cite{Freidel:2022ryr} (see also\cite{Berglund:2022qsb})
combining the properties of a quantized phase space with the Bekenstein (holographic) bound in a gravitational setting\cite{tHooft:1993dmi},
followed by its realization in QFT (\SS\,\ref{s:QFT2}), and also in string theory (\SS\,\ref{s:mStr}).
In Section~\ref{s:Higgs}, we discuss the recently derived seesaw formula for the Higgs mass, and address the fermion mass and mixing structure in Section~\ref{s:ferMass}:
We introduce a cascade of analogous seesaw formulae also for Standard Model fermions (\SS\,\ref{s:Generalia}), resulting in a realistic pattern of their masses (\SS\,\ref{s:fMasses}) and mixing (\SS\,\ref{s:fMixing}).
All these results should be understood as bounds on the cosmological constant, the Higgs mass as well as the masses and mixing of quarks and leptons.
Our closing comments are collected in Section~\ref{s:Coda}.

\section{The Cosmological Constant}
\label{s:CC}

\subsection{The Problem}
\label{s:CCP}
We start our presentation with a discussion of the canonical
calculations of the cosmological constant in quantum field theory and in string theory. In particular, we emphasize the remarkable similarities (as well as crucial differences) between these two approaches.
\paragraph{Quantum Field Theory:}
Let us start with the QFT vacuum partition function (of a free scalar in $D$ 
spacetime dimensions, even though our discussion can be generalized for other fields)
\begin{equation}
    Z_{\text{vac}} 
    = \int\rD\phi~ e^{-\int \frac{1}{2}\phi (-\partial^2 + m^2) \phi} 
    \propto \frac1{\sqrt{\det(-\partial^2 + m^2)}},
\end{equation}
which we can rewrite as
\begin{equation}
    Z_{\text{vac}} =  e^{-\frac{1}{2}{\rm Tr\, log}(-\partial^2 + m^2)}.
\end{equation}
In momentum space, $-\partial^2 = k^2$, and also
\begin{equation}
    -\frac{1}{2} \log(k^2+m^2) = \int \frac{\rd \tau}{2 \tau} e^{-(k^2+m^2)\tau/2},
\end{equation}
where the Schwinger parameter $\tau$ is a worldline affine parameter (world-line time) associated with a particle (quantum of the field $\phi$).
 After taking the trace, we have
\begin{equation}
    \int \frac{\rd^D k}{(2\pi)^D}\log(k^2 + m^2) 
    = \int \frac{\rd^{D-1}k}{(2\pi)^{D-1}} \frac{\omega_k}{2},
\end{equation}
since
\begin{equation}
    \int \frac{\rd \tau}{2 \tau}\int \frac{\rd k^0}{2\pi}e^{-(k^2+m^2) \tau/2} = \frac{\omega_k}{2},
\end{equation}
where $\omega_k^2 = k^2+m^2$ with $\omega_k$ equivalent to $k_0$ on-shell.
Thus, the vacuum energy density in $D$ spacetime dimensions becomes
\begin{equation}
    \rho_0 = \int \frac{\rd^{D-1}k}{(2\pi)^{D-1}} \frac{\omega_k}{2} \sim \Lambda_D,
\end{equation}
with $\Lambda_D$ the volume of energy-momentum space.
 This is a divergent expression that leads to the cosmological constant problem (see also \cite{Donoghue:2020hoh}).\footnote{Using Einstein's equations the cosmological constant in $D=4$ dimensions is $\Lambda_{cc}\sim\rho_0 G_N \sim \rho_0 l_P^2$.}
Notice that the vacuum partition function is also 
\begin{equation}
        Z_{\text{vac}} =\bra{0}|e^{-iH t}|{0}\ket = e^{-i\rho_0 V_D},
\end{equation}
where $V_D$ is the volume of $D$-dimensional spacetime, and $\rho_0$ is the vacuum energy density.
Furthermore, $Z_{\text{vac}}=\exp\{Z_{S^1}\}$ where $Z_{S^1}$ is the partition function on $S^1$ in the world-line formulation
\begin{equation}
    Z_{S^1} = V_D \int\frac{\rd^D k}{(2\pi)^D} \int \frac{\rd \tau}{2 \tau}~
    e^{-(k^2+m^2)\frac{\tau}{2}} \define \int \frac{\rd \tau}{2 \tau} Z_{S^1}(\tau).
 \label{e:Z1}
\end{equation}
Thus, the vacuum energy density is given by (scaling as before)
\begin{equation}
        \rho_0 = \frac{iZ_{S^1}}{V_D} \sim \Lambda_D.
\end{equation}
This is an important expression that we will use repeatedly in what follows.

\paragraph{String Theory:}
For the case of a bosonic string, instead of one particle we have an infinite tower of particles with a stringy mass spectrum, as emphasized by Polchinski in \cite{Polchinski:1985zf,Polchinski:1998rq}
\begin{equation}
        m^2 = \frac{2}{\alpha'}(h+\bar h - 2).
\end{equation}
Thus, summing over the physical string states (``phys.\,st.'') we obtain
\begin{equation}
        \sum_{\rm phys.\,st.} Z_{S^1} = \sum_{h,\Bar{h}} V_D \int \frac{\rd r (2\pi r)^{-D/2}}{2 r}\int \frac{\rd \theta}{2\pi}e^{i(h - \bar h)\theta} e^{-\frac{2}{\alpha'}(h+\bar h -2)\frac{r}{2}},
\end{equation}
where we have imposed the level matching $h=\bar{h}$ (or $\delta_{h,\bar{h}}$), after integrating over $k$.
By defining $\tau=\theta + i \frac{r}{\alpha'}\define \tau_1 + i \tau_2$, as usual, the partition function of a bosonic string on $T^2$ is
\begin{equation}
        Z_{T^2} = V_D \int \frac{\rd \tau \rd \bar{\tau}}{2\tau_2} (4\pi^2 \alpha' \tau_2)^{-D/2} \sum_h q^{h-1} \bar{q}^{\bar{h}-1},
\end{equation}
where $q\define e^{2\pi i\tau}$. This can be derived directly from the Polyakov path integral.
Using $r\define\alpha'\tau_2$, we may rewrite:
\begin{equation}
        (4\pi^2\alpha'\tau_2)^{-D/2} =\int\frac{\rd^Dk}{(2\pi)^D}e^{-k^2\frac{r}{2}}.
\end{equation}
Thereby, just as in QFT,
\begin{subequations}
 \label{e:Z2}
\begin{equation}
    Z_{T^2} \define V_D \int\frac{\rd^D k}{(2\pi)^D} f(k^2) \define \int \frac{\rd \tau \rd \bar{\tau}}{2\tau_2} Z_{T^2} (\tau)\sim V_D\, \Lambda_D,
\end{equation}
with 
\begin{equation}
    \Lambda_D \define \int\frac{\rd^D k}{(2\pi)^D}
    \quad\text{and}\quad 
    f(k^2) \define \int_F \frac{\rd^2\tau}{2\tau_2} e^{-k^2 \alpha' \tau_2/2} \sum_h q^{h-1} \bar{q}^{\bar{h}-1},
\end{equation}
\end{subequations}
where $F$ is the fundamental domain. 
Note that $f(k^2)$ is dimensionless, so it does not contribute to the scaling of $Z_{T^2}$ and 
the vacuum energy $\rho_0 \sim Z_{T^2}/V_D$.
The only (crucial) difference is that the QFT region of integration is 
\begin{equation}
        |\tau_1|<\tfrac{1}{2},\quad\tau_2>0,
\end{equation}
whereas modular invariance of string theory reduces this to
\begin{equation}
        |\tau_1|<\tfrac{1}{2},\quad |\tau|>1.
\end{equation}
Thereby, the cosmological constant becomes {\em UV-finite\/} in string theory, but is still related to $\rho_0\sim Z_{T^2}/V_D\sim \Lambda_D$, so that the ``cosmological constant problem'' persists in a manner very similar to what we have already encountered for particles/quantum fields.
Note that supersymmetry (SuSy) does not help with this fundamental problem.
It is well known that unbroken SuSy implies flat/Minkowski or anti~de~Sitter ($\L_D<0$) spacetime, due to the cancellation of the bosonic and fermionic contributions, without changing the offensive phase space term in the above expression for the vacuum energy. Broken SuSy does allow for a positive cosmological constant, but the relationship $\rho_0 \sim \Lambda_D$ continues to hold.
In particular, SuSy (whether unbroken or broken) does not affect the crucial spacetime and momentum space volumes which are the ultimate cause of the cosmological constant problem.

\subsection{Resolving the Problem}
\label{s:CC-fix}
Given these insights regarding the vacuum energy and volumes of spacetime and momentum space we now present a way to deal with the cosmological constant problem.
 Following\cite{Freidel:2022ryr} (see also\cite{Berglund:2022qsb,Freidel:2023ytq}) we return to $Z_{S^1}$ as given in~\eqref{e:Z1}, and for simplicity we set $m=0$ and with $p$ denoting the momentum,
 even though our discussion is valid for $m \neq 0$. (Also our discussion is valid for other,
 and not only scalar, fields,
 as well as the one-loop string partition function $Z_{T^2}$.)
 So
\begin{equation}
    Z_{S^1} = V_D \int \frac{\rd \tau}{2\tau}
    \int\frac{\rd^Dp}{(2\pi)^D}~e^{-\frac{p^2 \tau}{2}},
\end{equation}
where the spacetime volume is $V_D = \int \rd^D q$. This naturally leads to the {\em phase space} expression
\begin{equation}
     Z_{S^1} = \int\frac{\rd \tau}{2\tau}~ Z(\tau),\quad
    Z(\tau) = \int\frac{\rd^D q}{(2\pi)^D}\int \rd^D p~ e^{-\frac{p^2\tau}{2}} \define \Tr e^{-\frac{p^2\tau}{2}},
\end{equation}
where ${\Tr}$ is now defined over the {\em phase space}, and
we postpone the $\tau$-integration until the very last step.

\paragraph{Phase Space:}
In $D=4$ we can regularize the above phase space expression as follows
\begin{equation}
    Z(\tau) = \prod_{i=1}^4 \frac{1}{2\pi}\int_{-\infty}^\infty \rd q_i \int_{-\infty}^\infty \rd p_i~ e^{-\frac{p_i^2\tau}{2}},
 \label{e:Z(t)}
\end{equation}
which by discretizing phase space may be written as
\begin{equation}
    Z(\tau) = \left(\frac{\lambda \,\epsilon}{2\pi} \sum_{k,\tilde{k}\in {\mathbb{Z}}} \int_0^1\rd  x\,\rd \tilde{x}~ e^{-\frac{(k+\tilde{x})^2\epsilon^2\tau}{2}} \right)^4,
 \label{e:Z(t)2}
\end{equation}
where we have redefined $p\to \epsilon\,\tilde{x}$, $q\to \lambda\,x$,
with the two scales related by\,\footnote{\label{fn:chbar}As usual, we omit writing explicit factors of  $\hbar$. They are indicated here explicitly to emphasize that $\epsilon$ and $\lambda$ are a {\em momentum\/}- and {\em length\/}-scales, respectively.} $\lambda\,\epsilon= \hbar$.
While the expression~\eqref{e:Z(t)2} is still divergent, restricting the sum to finite range leads to the
 {\em modular regularization}\cite{Freidel:2015pka,Freidel:2015uug}:
\begin{equation}
    Z(\tau) = \left(\frac{\lambda \,\epsilon}{2\pi} \sum_{k=0}^{N_q-1} \sum_{\tilde{k}=0}^{N_p-1} \int_0^1\rd x\,\rd \tilde{x}~
    e^{-\frac{(k+\tilde{x})^2\epsilon^2\tau}{2}} \right)^4.
 \end{equation}
Here, $N_q,N_p$ count the number of unit cells\footnote{From the point of view of modular polarization\cite{Freidel:2016pls} this counts unit cells of vacua, rather than on-shell states or particles.} in the spacetime and momentum space dimensions, respectively.
In turn, this expression prompts the definitions:
\begin{equation}
     l\define N_q \lambda,\quad \text{and}\quad
     M_\L \define N_p \epsilon,
     \qquad\text{with}\quad N=(N_p N_q)^4\in \mathbb{Z},
 \label{e:N}
\end{equation}
so that $l^4\define V_4$ is the size (4-volume) of spacetime, $M_\L^4$ is the size (4-volume) of energy-momentum space.
Thus (see footnote~\ref{fn:chbar}),
\begin{equation}
    l^4 M_\L^4 = N\,\hbar^4,
    \quad \text{or}\quad
    M_\L^4 = \frac{N\,\hbar^4}{l^4},
 \label{e:lLN}
\end{equation}
but there is actually an upper bound on $\rho_0\sim M_\L^4\leq \frac{N}{l^4}$ in $D=4$ owing to the fact that $\exp\{-p^2\tau/2\}\leq 1$ in~\eqref{e:Z(t)}.
The same bound also holds in string theory, following our earlier calculation of the partition function of the bosonic string on $T^2$ in $D=4$:
\begin{equation}
    \rho_0 \leq \frac{N}{l^4} .
 \label{e:rho-bound}
\end{equation}
This reasoning can be extended for quantum fields and effective potentials in QFT, including the cosmological phase transitions (electroweak and QCD), without changing the outcome for this bound on the vacuum energy, as we will explain in the following section.

\paragraph{Holography:}
Given the above positive cosmological constant, consider now the Bekenstein bound in a four-dimensional space with a cosmological horizon, in other words, the fact that we have a de~Sitter spacetime.
In static coordinates, de~Sitter spacetime metric is 
\begin{equation}
    \rd s^2_{dS} = - \Big(1-\frac{r^2}{r_{CH}^2}\Big) \rd t^2
    + \frac{\rd r^2}{\big(1-\frac{r^2}{r_{CH}^2}\big)}+r^2\rd \omega^2_{S^2},
\end{equation}
where $l\define r_{CH}$, the cosmological horizon, is the size of the observed spacetime.
By identifying the above microscopic counting of ground states with the gravitational entropy, following the discussion in\cite{Freidel:2022ryr,Freidel:2023ytq},  the Bekenstein bound  ($S_{\text{grav}} = l_P^{-2}\textit{Area}$) then becomes\cite{Bekenstein:1980jp}
\begin{equation}
    N\leq \frac{l^2}{l_P^2}.
 \label{e:NllP}
\end{equation}
Combining this bound with the bound~\eqref{e:rho-bound} on $\rho_0$ produces
\begin{equation}
    \rho_0\leq \frac{1}{l^2\, l_P^2},
  \label{e:rho=llP}
\end{equation}
which reveals a mixing of the UV ($l_P$) and the IR ($l$) scales.
This mixing in turn produces a {\em bound} for the cosmological constant in $D=4$ dimensions, $\Lambda_{cc}=\rho_0\, l_P^2$:
\begin{equation}
    \Lambda_{cc} \leq \frac{1}{l^2}.
\end{equation}
The natural energy scale, $\epsilon_{cc}$, associated with the vacuum energy density, is then
\begin{equation}
    \rho_0 = \epsilon_{cc}^4\sim \frac{1}{l^2\, l_P^2},
\end{equation}
with a corresponding natural length scale, $l_{cc}\simeq 1/\epsilon_{cc}$, given by the the seesaw formula
\begin{equation}
    l_{cc}\simeq \sqrt{l\, l_P},
    \qquad\text{i.e.,}\qquad
    M_\L\simeq\sqrt{M\,M_P},
  \label{e:lcc=llP}
\end{equation}
where $M\sim10^{-34}$\,eV is the Hubble mass-scale and $M_P \sim 10^{19}$\,GeV, is the Planck scale.
Finally, note that the integration over the world-line or the world-sheet parameters does not change this final result. These integrations should be done at the end, and they only provide an overall renormalization of the Newton constant.

The above results have the following noteworthy properties:
\begin{itemize}\itemsep=-3pt\vspace*{-1mm}
  \item With $l\sim 10^{27\text{--}28}$\,m and $l_P \sim 10^{-35}$\,m, the seesaw relation~\eqref{e:lcc=llP} yields
    $l_{cc}\simeq 10^{-4}$\,m or $\epsilon_{cc}\simeq 10^{-3}$\,eV,
    in agreement with observations and identifiable as $\epsilon_{cc}=M_\L$ from~\eqref{e:N}.
    \item The relation~\eqref{e:rho=llP} naturally gives $\rho_0\to 0$ when $l\to\infty$, and $l$ is the IR length-scale.
    \item The relation~\eqref{e:lcc=llP} is radiatively stable since there is no UV dependence.
    \item Thus, essentially, the cosmological constant is small because the universe is filled with a large number of degrees of freedom: $N \sim 10^{124}$.
    \item In turn, $N$ is large because fluctuations scale as $\frac{1}{\sqrt{N}}$ and are small, indicating the stability of the universe.
    \item This estimates $N_i$ (where $i$ is $t,x,y,z$) as $N^{1/4} \sim 10^{31}$, which is not so unreasonable in comparison with Avogadro's number, $10^{23}$, for matter degrees of freedom.
\end{itemize}
We emphasize furthermore the quantum contextuality of the above calculation: The measurement of a quantum observable depends on which commuting set of observables are within the same measurement set of observables, 
i.e., quantum measurements depend on the {\em context\/} of measurement.
\begin{itemize}\itemsep=-3pt\vspace*{-1mm}
    \item First, $\epsilon$ is {\em not} a cut-off, since $\epsilon$ and $\lambda$ can be arbitrary, albeit related by $\lambda \,\epsilon =\hbar$.

    \item Second, $\epsilon^4$ is effectively eliminated in favor of $N$, which is the new quantum number, and the size of spacetime, $l=r_{CH}$, the cosmological horizon, i.e., the size of the observed classical spacetime.

    \item $N$ is determined by the Bekenstein bound,~\eqref{e:NllP} and is thereby related to $l$ and $l_P$ (the ultimate IR and UV scales, respectively), which is where gravity enters, 
    via $G_N\sim l_P^{2}$.

\end{itemize}
By contrast, effective field theory cannot ``see'' $N$, and in particular does not ``know'' about the Bekenstein bound or the UV/IR mixing.
For example, vacuum energy routinely cancels in the computation of EFT correlation functions. Also, EFT {\em is defined\/} in classical spacetime\cite{Weinberg:1974tw,Weinberg:1978kz}\footnote{The above seesaw relation  $l_{cc}\simeq \sqrt{l\, l_P}$ does appear in \cite{Cohen:1998zx},
where $l$ and $l_P$ are, respectively, the ultimate IR and UV length-scales in EFT, and are related by the physics of black holes/holographic bound. However, there one does not have the crucial aspects of our derivation: modular representation, the number of phase space cells $N$, the explicit UV/IR mixing and contextuality. We will comment on the connection with EFT at the end of Section~\ref{s:QFT2}.}.
Thus the above calculation calls for a fundamental quantum formulation that relies on the modular polarization\cite{Freidel:2016pls}. Precisely this is provided by the metastring formulation of string theory\cite{Freidel:2015pka},
which we will duly discuss in Section~\ref{s:mStr}.

\subsection{The Cosmological Constant in QFT and Phase Space}
\label{s:QFT2}
Having summarized our recent work, we now turn to discuss the textbook QFT computation of the vacuum energy and illustrate how the reasoning in the previous sections applies in that context. In particular, we show that the above argument is robust even in the light
of realistic cosmological phase transitions.
Essentially, the phase space argument combined with the Bekenstein bound
that leads to the above bound for the cosmological constant~\eqref{e:lcc=llP}
is globally valid and is insensitive to any 
EFT redefinition of the QFT vacuum due to possible phase transitions.

Let us consider a scalar field theory\cite{rSW1,rAZ-QFT},
knowing that the following computation can be generalized for fermions and other fields (and it could be used in string theory as well). The relevant Lagrangian is
\be
L = \frac{1}{2} (\partial \phi)^2  - \frac{1}{2} m^2 \phi^2 - \frac{1}{24}g \phi^4 \define \frac{1}{2} (\partial \phi)^2 - V(\phi).
\ee
The partition function is defined as
\be
Z(J) \define e^{i W(J)} = \int \rD \phi~ e^{i [S(\phi) + J\phi]},
\ee
where $W(J)$ is the generating functional of vacuum correlation functions, and it represents a direct analogue of
the partition function for a particle on a circle, or a string on a torus from the previous section.
Given $W(J)$, we can define it's Legendre transform to obtain
$\Gamma(\phi)$, the effective action, as
\be
\Gamma(\phi) \define W(J) - \int\rd^4x~ J(x) \phi(x).
\ee
The leading term in the expansion of $\Gamma(\phi)$ is the
effective potential,
\be
\Gamma(\phi) \define \int\rd^4 x [-V_{\text{eff}}(\phi) +\dots],
\ee
the minimum of which defines the vacuum energy in QFT.
(The proper normalization of the path integral that is responsible for the vacuum energy is here absorbed in the expression for the effective potential $V_{\text{eff}}$.)

To determine the $\hbar$-expansion of this expression (which is the source of problems with vacuum energy in QFT), one expands the original action $S$ around its classical minimum and looks at the leading quadratic fluctuations. Upon the evaluation of the relevant Gaussian integral one obtains the generating functional $W(J)$ or alternatively, its Legendre transform, the effective action,
\be
\Gamma(\phi) = S(\phi) +\frac{i \hbar}{2}\,\Tr\log\big[\partial^2 +V''(\phi)\big],
+ O(\hbar^2).
\ee
Thus, the effective potential $V$ reads as follows
\be
V_{\text{eff}}(\phi) = V(\phi) - \frac{i \hbar}{2}\int \frac{\rd^4 k}{(2 \pi)^4}\log[{k^2 -V''(\phi)}]+ O(\hbar^2).
\ee
An explicit evaluation of this momentum integral gives the
famous Coleman-Weinberg potential\cite{Coleman:1973jx}.
Using the Schwinger parametrization of the 
logarithm
\be
-\frac{1}{2}\log\big[ U(k^2, \phi)\big]
= \int \frac{\rd r}{r}~e^{-U(k^2, \phi)\,r/2},
\ee
leads to the following crucial observation:
The quantum corrections to the effective action $\Gamma$ are
given by an integral over phase space (as in 
the computation summarized in the previous section)
\be
\Gamma (\phi) \sim \int \rd^4 x \int \frac{\rd^4 k}{(2 \pi)^4}
\int \frac{\rd r}{r}~ e^{-U(k^2, \phi)\,r/2},
\ee
where the exponent in the above integral is bounded by 1.
This is completely analogous to the expression for the partition function of a particle on a circle~\eqref{e:Z1}, or a string on a torus.
Thus, after using modular regularization, we are led to the same bound on the vacuum energy evaluated from the effective action, when the QFT is coupled to gravity and subjected to the Bekenstein bound. (In this reasoning the integration of the Schwinger parameter is done at the end, and its effect is absorbed in the renormalization of the Newton constant.)

The canonical reasoning tells us that $\Gamma(\phi)$, evaluated at the minimum of the effective potential, does give a divergent expression for the vacuum energy, which upon coupling of this simple field theory to gravity leads to the cosmological constant problem.
Such a cosmological constant is very sensitive to radiative corrections, to the existence of phase transitions as captured by the effective potential, etc.
In particular, at finite temperature $T$, the Landau-Ginsburg effective Lagrangian may be written as
\be
L = \frac{1}{2} (\partial \phi)^2  - \frac{1}{2} a(T-T_c) \phi^2 - \frac{1}{24}g \phi^4 \define \frac{1}{2} (\partial \phi)^2 - V(\phi, T),
\ee
where $T_c$ is the critical temperature, and $a >0$ and $g >0$.
Above $T_c$ this has a global minimum, and below the critical temperature this describes symmetry breaking with an unstable (tachyonic) local maximum and a new global minimum determined by the expectation value of the order parameter, $\phi$, --- the Higgs field, upon the inclusion of the gauge field in this description.
However, we can repeat the above calculation of the effective action at finite temperature $T$ and deduce that the quantum part of the effective action again scales as
\be
\Gamma (\phi, T) \sim \int\rd^4 x \int \frac{\rd^4 k}{(2 \pi)^4}
\int \frac{\rd r}{r} e^{-U(k^2, \phi, T)\,r/2},
\ee
where the exponent is yet again bounded by 1, given the generic
positive definite nature of $U(k^2, \phi, T)$.
Therefore, the already established bound for the vacuum energy (determined by the minimum of this finite temperature effective potential) is valid. This, when combined with modular regularization and the Bekenstein bound, gives the same seesaw formula~\eqref{e:lcc=llP}.
Even more explicitly, at finite temperature $T$ 
(with $\beta \define 1/T$), we have
\be
\int (\,\cdots) \frac{\rd^4 k}{(2 \pi)^4} \to 
T \sum_{k^0=2 \pi i n T} \int (\,\cdots) \frac{\rd^3 k}{(2 \pi)^3},
 \label{e:k0->T}
\ee
by the usual rules of QFT at finite temperature.
Therefore, the finite temperature addition to the effective potential
reads as follows
\be
V_{\text{eff}} (T) \sim  \frac{T}{2}\sum_n \int \frac{d^3 k}{(2 \pi)^3}\log[{4\pi^2 n^2 T^2 +{\vec{k}}^2 +V''(\phi)}],
\ee
or by using the Schwinger parametrization
\be
\Gamma (\phi, T) \sim T  \sum_n\int \rd^4 x \int \frac{\rd^3 k}{(2 \pi)^3}
\int \frac{\rd r}{r}~ e^{-[\,4\pi^2 n^2 T^2 + U({\vec{k}}^2, \phi, T)\,]\,r/2}.
\ee
Note that $\Gamma (\phi, T)$
is still bounded by the volume of phase space; 
since $T$ measures the size of the ``imaginary time/energy'' direction and the indicated summation stems from having discretized that direction in~\eqref{e:k0->T}. Finally, returning to the continuum then recovers the expected $\sum_n\int\rd^3k\to\int\rd^4k$. 

In all of these calculations the integration of the Schwinger parameter $r$ is done at the end of the calculation, where it contributes only to the renormalization of the Newton constant, without any influence on the bound on the vacuum energy.
The effective action (before the $r$ integration) is bounded by the phase space volume, which when combined with modular regularization and the Bekenstein bound, gives the already derived bound on the cosmological constant.

Therefore, the above expressions for the effective action and the evident appearance of the volume of phase space justifies applying our argument from the previous section, which together with modular regularization and the Bekenstein bound leads to the same bound on the cosmological constant in the context of a QFT coupled to gravity and reproduces the seesaw formula~\eqref{e:lcc=llP} for the vacuum energy in QFT.
Given the nature of the bound and the ensuing seesaw relation for the cosmological constant, this bound is radiatively stable and it holds true even after the cosmological phase transitions (electroweak, QCD) happen.
The reason for this is simply that the bound stems from the mixing~\eqref{e:rho=llP} of the UV (gravitational, Planck, scale) and the IR scale (the size of the observed universe, the Hubble scale).
 Such global (non-local) features associated
with modular regularization of phase space are inherently absent in local QFT, as well as any EFT, which by construction ``sees'' neither this UV/IR mixing nor the ensuing resolution of the vacuum energy problem.

Let us emphasize that in the context of EFT the vacuum energy is associated with the normalization of the path integral, and thus it cancels in the usual EFT calculations (without gravity). In our recent review\cite{Berglund:2022qsb}
we highlight how the canonical EFT results emerge in a double scaling limit in which the new quantum number $N \to \infty$,
as well as $l \to \infty$ with 
the constant ratio $N/l^4 \define 1/l_{\Lambda}^4 = \textit{finite}$, while
$l_P \to 0$. The seesaw nature of the 
formula for $l_{\Lambda}$ is however preserved in this
doubled scaling limit, leading to the observation of\cite{Cohen:1998zx} (where $l$ is interpreted as the EFT's IR cut-off, and $l_P$ as its UV cut-off).
However, EFT is defined in classical (and not quantum) spacetime, and it is fundamentally insensitive to the UV/IR mixing, whereas quantum gravity crucially 
depends on 
quantum spacetime degrees of freedom and the mixing between the UV and IR physics.

\subsection{Realization in String Theory} 
\label{s:mStr}
As already emphasized in Section~\ref{s:CC-fix}, the above phase space/modular formulation is naturally realized in terms of a chiral, phase-space-like and T-duality covariant reformulation of the bosonic string, the {\em metastring}\cite{Freidel:2013zga,Freidel:2015uug,Freidel:2015pka} (which may also be turned into a non-perturbative proposal\cite{rBHM10,Berglund:2021hbo,Berglund:2022qsb}):
\begin{equation}
S^{\text{ch}}_{\text{str}}=
\int
\rd \tau\,\rd \sigma~
    \Big[\pa_{\tau}{\X}^{a} \big(\eta_{ab}(\X)+\omega_{ab}(\X)\big)
    -\partial_\sigma\X^a H_{\!ab}(\X)\Big] \partial_\sigma\X^b. 
\end{equation}
Here, $\X^a\define (X^a/\ls ,\tilde X_a/\ls )^{T}$ are
coordinates on phase-space like (doubled) target spacetime and the fields $\eta, H,\omega$ are all dynamical (i.e., generally $\X$-dependent) target spacetime fields. In terms of the left- and right-moving 0-modes of the bosonic string, one defines
\begin{equation}
    x^a\define x^a_L + x^a_R,\quad \Tilde{x}^a \define \Tilde{x}^a_L - \Tilde{x}^a_R.
\end{equation}
In the context of a {\em flat\/} metastring, the coefficients
 $\eta_{ab}$,  $H_{ab}$
 and $\omega_{ab}$\footnote{Setting $\omega_{ab}=0$ directly connects $S^{\text{ch}}_{\text{str}}$ to double field theory.} are constant:
\be\label{etaH0} 
	\eta_{ab} = \left( \begin{array}{cc} 0 & \delta \\ \delta^{T}& 0  \end{array} \right),\quad
H_{ab} =  \left( \begin{array}{cc} h & 0 \\ 0 &  h^{-1}  \end{array} \right),
\quad \omega_{ab} = \left( \begin{array}{cc} 0 & \delta \\ -\delta^{T}& 0  \end{array} \right) ,
\ee
where $h$ denotes the flat $(1,d{-}1)$-dimensional metric and $\d$ is the Kronecker symbol.
The standard Polyakov action is then obtained by setting $\omega_{ab}=0$ and integrating out the $\tilde x_a$,
\begin{equation}
    S_P=\int \rd \tau\,\rd \sigma~ \gamma^{\alpha\beta}\, \partial_\alpha X^a\, \partial_\beta X^b\, h_{ab} + \ldots
\end{equation}
The triplet $(\omega,\eta,H)$ defines Born geometry\cite{Freidel:2013zga,Freidel:2015pka} (which is ultimately dynamic, suggesting a ``gravitization of quantum theory''\cite{Freidel:2014qna,Berglund:2022qcc,Berglund:2022skk}) so that the metastring  propagates in a (dynamical) modular spacetime, a phase space like structure that naturally arises in any quantum theory\cite{Freidel:2016pls}. One of the key consequences of this is that the metastring is intrinsically non-commutative and also that its low energy QFT-like description in modular spacetime is intrinsically non-commutative.
Thus every Standard Model field $\phi(x)$ is doubled as $\phi(x, \tilde{x})$ and $\tilde{\phi}(x, \tilde{x})$,
with doubled and non-commutative arguments
$[x^a, {\tilde{x}}_b] = i\ls^2\, \delta^{a}{}_{b}$.
The quanta of such modular fields are the zero modes of the metastring --- the {\em metaparticles} --- whose dynamics are given by a world-line action involving a doubling of the usual phase space coordinates. The metaparticle (``mp'') action is of the form
\cite{Freidel:2017wst, Freidel:2017nhg, 
Freidel:2018apz, Freidel:2021wpl}
\begin{equation}\label{mp1}
S_{\text{mp}} \define \int_0^1 d\tau \Big[p\cdot \dot x +\tilde p\cdot \dot{\tilde x}+ \alpha'\, p \cdot\dot{\tilde p}
- \frac{N}2\left(p^2 +{\tilde p}^2 + \mathfrak{m}^2\right) +{\tilde N}\left(p\cdot \tilde p - \mu \right)\Big]\,,
\end{equation}
where the dot-product denotes contraction with signature $(-,+,\ldots,+)$.
The new feature here is the presence of a non-trivial symplectic form on the metaparticle phase space, the non-zero Poisson brackets being
\beq
\{p_\mu, x^\nu\}=\delta_\mu^\nu,\,\,\,
\{\tilde p_\mu, \tilde x^\nu\}=\delta_\mu^\nu,\,\, \,
\{ \tilde x_\mu,x^\nu \}= \pi\alpha'\,\delta_\mu^\nu,\quad
\alpha'\sim\ls^2.
\label{xtxcomm}
\eeqn
with $\mu,\nu=0,1,\dots,d-1$.
Because of its interpretation as a particle model on Born geometry, associated with the modular representation of quantum theory,
the space-time on which the metaparticle propagates is ambiguous, with different choices related by what in string theory we would call T-duality.
The attractive features of this model include world-line causality and unitarity, as well as an explicit mixing of widely separated energy-momentum scales.

In the generic modular polarization of quantum theory, instead of considering the standard commutation relations between the position and momentum operators, one considers the generators of translations in {\em phase space}
\begin{equation}
    \hat U_a = e^{\frac{i}{\hbar} \hat p \,a},\quad
     \hat V_{\frac{2\pi \hbar}{a}} = e^{\frac{i}{\hbar} \hat q \frac{2\pi \hbar}{a}}\quad  \implies
     [\hat U_a, \hat V_{\frac{2\pi \hbar}{a}}] =0.
\end{equation}
In terms of modular variables introduced by Aharonov and collaborators\cite{Aharonov:2005uc},
\begin{equation}
    [\hat q]_a \define \hat q\, {\rm mod}\, a\quad
    [\hat p]_\frac{2\pi \hbar}{a} \define \hat p\, {\rm mod}\, \frac{2\pi \hbar}{a}\quad \implies
    \big[\,[\hat q]_a, [\hat p]_\frac{2\pi \hbar}{a}\,\big]=0.
\end{equation}
The {\em space} of commuting subalgebras of the Heisenberg algebra, $[\hat q,\hat p]=i\hbar$, which in the covariant (self-dual lattice) phase space formulation becomes the modular spacetime\cite{Freidel:2015pka,Freidel:2015uug,Freidel:2016pls} {\em is\/} the target space of the metastring\cite{Freidel:2015pka} and 
the metaparticles \cite{Freidel:2017wst, Freidel:2017nhg, Freidel:2018apz}.
Vertex operators in metastring theory are representations of this Heisenberg algebra. This description (intrinsically non-commutative, since $[x,\tilde x]=2\pi i\ls^2$) however avoids all of the co-cycles that turn up in standard descriptions of 
the vertex operator algebra in string theory\cite{Polchinski:1998rq,Polchinski:1998rr}.

A more elementary (and familiar) argument for the existence of modular spacetime may be presented as follows: In quantum theory, short (UV) distances are associated with high energy, as implied by the indeterminacy relation, $\delta q \sim 1/\delta p$ (in $\hbar=1$ units). On the other hand, in classical (as well as semiclassical) gravity, the Schwarzschild radius $R_S$ of a mass $M$ is given by $R_S \sim G M \sim l_P^2 M$,
where $G \sim l_P^2$ is the gravitational constant in 4-dimensional spacetime, with $l_P$ the Planck length.
In quantum gravity, quite generally, one therefore expects that higher energy leads to larger (IR) distances $\delta q\sim l_P^2\,\delta p$.
These diametrically contrasting behaviors (associated with UV and IR) may be reconciled by relating the UV and IR physics:
Recall that, given a fundamental lattice length, quantum states are described in terms of quantum numbers associated with both a lattice and its dual\cite{Freidel:2016pls}. In our present case, this involves momenta $p$ and their duals $\tilde{p}$, provided that these commute $[p, \tilde{p}]=0$.
The indeterminacies $\delta p$ and $\delta\tilde{p}$ thereby being interchangeable provides the first substitution in the chain:
 $l_P^2\,\delta p \sim \delta q \to l_P^2\,\delta \tilde{p} \sim \delta q
  \to l_P^2 (\delta \tilde{q})^{-1} \sim \delta q
  \Rightarrow \delta q\,\delta \tilde{q} \sim l_P^2$,
where the second replacement used the canonical
$\delta\tilde{p}\,\delta\tilde{q}\sim1$ indeterminacy relation.
This implies a new fundamental non-commutativity between spacetime and dual spacetime coordinates  $[q, \tilde{q}] \sim i l_P^2$.
The commutative nature of modular variables in quantum theory insures that this can be completely 
covariantized\cite{Freidel:2016pls}.
Thus, combining the fundamental quantum and gravitational relations between spatial distances and momenta  leads to:
\begin{itemize}
\itemsep=-3pt\vspace*{-1mm}
\item 
 the concept of dual spacetime,
\item 
 the fundamental non-commutativity between spacetime and dual spacetime,
 \item 
 the Heisenberg algebras:
$[q, \tilde{q}] = i l_P^2$,
$[q, p] = i$,
$[\tilde{q}, \tilde{p}] = i$,
$[p, \tilde{p}] =0$.
\end{itemize}

Precisely this algebraic structure is realized in metastring theory\cite{Freidel:2013zga,Freidel:2015uug,Freidel:2015pka}, merely with ``softening'' the indeterminacy by replacing, $l_P\to\ls$, the length-scale in~\eqref{xtxcomm}.
The central point here is that the metastring formulation explicitly realizes modular spacetime and the modular polarization needed in the argument for the bound of the cosmological constant, and, as a theory of quantum gravity, also realizes the Bekenstein bound.
Thus, a natural realization of the resolution of the cosmological constant problem is found in the metastring formulation of string theory.
Moreover, string theory is a quantum theory of gravity {\em and\/} Standard Model-like matter. Therefore, other vexing problems beside the cosmological constant problem should be possible to address in the same context, to wit, the gauge hierarchy problem and the problem of fermion masses. 

\section{The Cosmological Constant and the Higgs Mass} 
\label{s:Higgs}

In order to address a stringy bound on the Higgs mass and the gauge hierarchy problem we concentrate on the new feature in the metastring formulation of the bosonic string captured by the
$[x,\tilde x]=2\pi i\ls^2$ non-commutativity, implying a new  Heisenberg algebra\cite{Freidel:2017wst,Freidel:2017nhg}  
\begin{equation}
    [\X^a,\X^b]=  i \omega^{ab}\, \ls^2 \implies [X^a,\Tilde{X}^b]
    =i\delta^{ab} \ls^2  \label{e:XtwX}
\end{equation}
in addition to the standard ones, with $\Pi_b, \Tilde{\Pi}_b$, the respective conjugate momenta to $X^a,\Tilde{X}^a$,
\begin{equation}
    [\X^a,{\rm I}\!\Pi_b]=  i \hbar \delta^a_b \implies [X^a,\Pi_b]=i\delta^a_b\hbar,\quad 
    [\Tilde{X}^a,\Tilde{\Pi}_b]=i\delta^a_b\hbar.  \label{e:XP+tw}
\end{equation}
We note that, if the Kalb-Ramond $B_{ab}$ (the source of the axion in four spacetime dimension) is constant and non-zero, dual coordinates do not commute. 
In general, for dynamical backgrounds, one has to deal with intrinsic 
non-associativity\cite{Freidel:2017nhg}. 
Note that the zero modes of the metastring --- { metaparticles\/}\cite{Freidel:2018apz,Barnes:2021akh} 
(which look like little rigid strings, or correlated pairs of particles and their duals), inherit these generic properties of the metastring\footnote{In what follows, we concentrate on the 4 spacetime dimensional $\ls$, being aware of different effective values for $\ls$ in 4 as opposed to 10 or any
other spacetime dimension.}; each Standard Model particle has a correlated ``dual,'' a candidate dark matter companion\cite{Berglund:2020qcu,Berglund:2021xlm}.

\paragraph{Cosmological Scale:}
We now repeat the above argument for the vacuum energy based on the modular regularization of the phase space volume, combined with the Bekenstein bound\cite{Bekenstein:1980jp}, but for the modular space of the Heisenberg algebra~\eqref{e:XtwX} in the context of the metastring formulation of string theory.
We claim that this leads to a bound on the Higgs mass, analogous to the above bound on the cosmological constant. The most important point here is that the relevant IR scale is provided by the cosmological constant mass scale, --- since the Higgs field {\em determines\/} the vacuum.

In $D=4$-dimensional spacetime and with $N_{\Lambda}$ fluxes in the modular space of $X^a,\Tilde{X}^a$, the  length scales of the $X^a$ and $\Tilde{X}^a$ are $l_\Lambda$, associated with the vacuum energy, and $\Tilde{l}$, respectively. Therefore, 
\begin{equation}
    l^4_\Lambda\, \Tilde{l}^4 = N_{\Lambda}\, (\ls^2)^4,
 \label{e:lLlN}
\end{equation}
analogously to the relation~\eqref{e:lLN},
for the first of the Heisenberg algebras in~\eqref{e:XP+tw}.
The holographic bound\cite{tHooft:1993dmi,Susskind:1994vu} for the effective spacetime associated with the vacuum energy is  $N_{\Lambda}=l_\Lambda^2/l_P^2$, which with~\eqref{e:lLlN} produces
\begin{equation}
    l_\Lambda\,\Tilde{l}=\ls^2 \left(\frac{l_\Lambda^2}{l_P^2} \right)^{1/4} = 
    \ls^2 \left(\frac{l_\Lambda}{l_P} \right)^{1/2}.
\end{equation}
The string length $\ls$ and the Planck length $l_P$  are related via the string coupling $g_s$
\be
g_s\,\ls = l_P, \quad\text{i.e.}\quad M_s = g_s\,M_P,
\ee
where $M_P$ is the Planck (energy) scale and $M_s$ the string (energy) scale.
The dual spacetime scale is thereby determined:
\be
    \tilde {l} = \frac{\ls^2}{\sqrt{l_\Lambda\, l_P}}
    = \frac{l_P}{g_s^2} \Big(\frac{l_P}{l_{\Lambda}}\Big)^{\!1/2}.
\ee 
We also demand that the extent of the dual spacetime is  $l_P$, i.e., we require that $\tilde{l} \define l_P$, which is consistent with the
assumption that the size of the ``dual phase space'' (defined by the dual spacetime and dual momenta) is Planckian.
This implies that
\be
g_s = \Big(\frac{l_P}{l_{\Lambda}}\Big)^{\!1/4} \define\Big(\frac{M_\L}{M_P}\Big)^{1/4}
      \to g_s^2 = \Big(\frac{M_\L}{M_P}\Big)^{1/2}\ll 1,
\ee
where $M_\L$ is the energy scale associated with the vacuum energy. Since $g_s\ll 1$,
the one-loop computation of
the (metastring) partition function is indeed a good approximation.

\paragraph{Higgs Mass:}
We now apply these results to the recent stringy relation between
the Higgs mass and the cosmological constant derived by Abel and Dienes\cite{Abel:2021tyt},
\be
m_H^2 = \frac{\xi M_\L^4}{M_P^2} - \frac{g_s^2 M_s^2}{8\pi^2} \vev\cX,
\label{e:AbelDienes}
\ee
where $\cX$ is a suitably normalized insertion in the second moment of the partition function and $\xi$ a modular completion to the terms in $\cX$.
Even though this formula was obtained in the bosonic string theory by paying attention to stringy modular invariance, the same formula could be obtained in the metastring reformulation of the bosonic string.
In fact, with $M_s=g_sM_P$ and $g_s=(M_\L/M_P)^{1/4}$, 
the Abel-Dienes formula becomes\footnote{As the insertion vev turns out to be negative, $\vev\cX<0$, we will use $|\vev\cX|$.}
\begin{equation}
    m_H^2 = \xi\frac{M_\L^4}{M_P^2} + \frac{|\vev\cX|}{8\pi^2}g_s^4 M_P^2
    \,=\, \xi\frac{M_\L^4}{M_P^2} + \frac{|\vev\cX|}{8\pi^2}M_\L M_P,
\label{e:AbelDienes2}
\end{equation}
a linear combination of the ${\sim}\,(M_\L^2/M_P)^2$ (``seesaw-light'') and the ${\sim}(\sqrt{\,M_\L M_P})^2$ (geometric mean) terms, each reflecting a seesaw relation of two scales, $M_\L$ and $M_P$.
The first term ($\sim10^{-34}$\,eV) is negligible compared to the second one, so that
\be
m_H \sim g_s\, M_s \sqrt{\frac{|\vev\cX|}{8 \pi^2} }
 \,=\, g_s^2\, M_P \sqrt{\frac{|\vev\cX|}{8 \pi^2} }
 \,\sim\, \sqrt{M_\L M_P }  \sqrt{\frac{|\vev\cX|}{8 \pi^2} },
\label{e:mH-BHM}
\ee
recovers the seesaw formula for the Higgs mass in string theory\cite{Berglund:2022qsb}, 
where $M_\L$ (${\sim}10^{-3}$\,eV) is energy scale associated with the vacuum energy, and $M_P$ (${\sim}10^{19}$\,GeV) is the Planck energy.
The geometric mean of these two energy scales is a TeV-scale, and thus the observed Higgs mass
can be obtained if $|\vev\cX| \sim 10^{-1}$ --- which is consistent with Abel \& Dienes' results\cite{Abel:2021tyt}.
We emphasize that the above formula, \eqref{e:mH-BHM}, should be understood as a bound on the Higgs mass, in analogy with the result~\eqref{e:lcc=llP} for the cosmological constant.

\paragraph{Summary:}
Analogously to our foregoing arguments for cosmological constant~\eqref{e:lcc=llP}, we have shown that
 ({\small\bf1})~examining the modular spacetime and metastring formulation of string theory, 
 ({\small\bf2})~combining the $[x,\tilde{x}]\neq0$ non-commutativity and holography in $x$-space,
 and
 ({\small\bf3})~assuming that $\Tilde{x}$ is of the Planck length size, leads in string theory to a seesaw formula also for the Higgs mass~\eqref{e:mH-BHM}. 
Like the one for the vacuum energy, this formula represents a bound provided by the size of the phase space and the Bekenstein bound in which the effective length scale is 
associated with vacuum energy $l_{\Lambda}$.
In this calculation, the two Heisenberg algebras in the metastring formulation ($[x,p]$ and $[x,\tilde{x}]$) are mutually consistent.

Having applied this logic to the formula for the Higgs mass \`a la Abel \& Dienes\cite{Abel:2021tyt} (derived in canonical bosonic string theory with manifest stringy modular invariance, but also compatible with its metastring formulation), 
we have not only arrived at a stringy bound for the Higgs mass, but also at a completely stringy view of the hierarchy problem. This directly ties the hierarchy problem to the vacuum energy problem, whereby the resolution of both lies in the fundamental (modular) phase-space approach combined with a Bekenstein bound on the number of relevant degrees of freedom.
 This new and unified understanding of these two central hierarchy problems
naturally points to metastring theory, and (as we outline in the next section) it can also address the problem of fermion masses.

Let us however conclude this section by addressing the naturalness of the above values for $N$, relevant for {\em\/both\/} hierarchy problems: the cosmological constant and the Higgs mass.
Both in statistical physics and in QFT, it is well known how to sum over contributions of closed diagrams:
 Simple combinatorics ensures that this is an exponent of the partition function associated with a closed loop (handle, for strings).
 As pointed out in Section~\ref{s:CCP}, the QFT vacuum partition function is 
 $Z_{\text{vac}} = \exp(Z_{S^1})$, with $S^1$ the circle of a vacuum loop traced by a particle;
 in string theory, one just replaces $S^1\to T^2$\cite{Freidel:2022ryr}.
 For the case of {\em dynamical} Born geometry\cite{Freidel:2014qna} (a generic feature of quantum gravity in the metastring formulation), the usual path integral measure $e^{iS}$ should be effectively replaced  by $e^{e^{iS}}$ after summing over handles of a dynamical quantum geometry, where in the approximation of a dilute gas of handles we have taken that the
effective partition function is just the canonical one.
 Summing over handles in the foamy quantum space\cite{Berglund:2023vrm}, from the point of view of the canonical complex geometry of quantum theory, thereby yields an effective action which is essentially $e^{iS}$.
 In the Euclidean formulation, this implies that the effective action at some scale sensitive to gravity can be {\em exponentially removed} from the natural scale of Planck gravity, indicating that the Higgs scale may well be where effects of quantum gravity could be seen.\footnote{Indeed, see the large class of widely usable toy models\cite{rBHM1,rBHM7,Berglund:2020qcu,Berglund:2021xlm}.}
Essentially, we claim the naturalness of the hierarchy of scales between the Higgs and
the Planck scale ultimately to be a quantum gravity effect, associated with ``gravitizing the quantum''\cite{Freidel:2014qna,Berglund:2022qcc,Berglund:2022skk}.
 Thus, the effective value of $N$ (per spacetime direction) that features in both hierarchy problems, the cosmological constant problem and the problem of the Higgs mass, is indeed naturally expected to be of the order of the familiar Avogadro number, and it is only genuine in the context of quantum gravity (or gravitized quantum theory) and quantized spacetime.

\section{On the Masses and Mixing of Quarks and Leptons}
\label{s:ferMass}
\subsection{General Comments}
\label{s:Generalia}
The above cosmological constant computation relies on the computation of the one-loop partition function for a particle or a string, or equivalently, the effective action for the effective field theory under consideration.
In each of these cases, we have shown that the relevant expression for the partition function and thus the cosmological constant is bounded by the phase space volume, in its modular regularization.
The Bekenstein bound for the number of the phase space cells then explicitly leads to a seesaw (geometric mean) formula for the cosmological constant.
Next, we have extended this computation to the evaluation of the Higgs mass. In that case we have relied on the
expression~\eqref{e:AbelDienes2} obtained by Abel and Dienes for the Higgs mass\cite{Abel:2021tyt} (derived also from a stringy partition function) that relates this quantity to the cosmological constant in string theory 
(definitely not an EFT feature), in the weak string coupling regime. 

Thus, at least for weak string coupling, the same computational strategy applies to both the Higgs mass
and the cosmological constant, and leads to the (geometric mean) seesaw formula.
 They differ only in the contextual UV and IR scales in the respective computations.
 The UV scale is $M_P$ for both, while the IR scale is the relevant Hubble mass scale $M$ for the cosmological constant scale, $M_\L$~\eqref{e:lcc=llP}, which then serves as the IR scale for $m_H$ in~\eqref{e:mH-BHM}. This relation follows both from the Abel-Dienes formula~\eqref{e:AbelDienes2},
as well as because of the physical meaning of the Higgs field --- it {\em\/determines\/} the vacuum of the matter sector and is responsible for the masses of all elementary particles, except neutrinos.

It is then only reasonable to ask: Should and does this reasoning extend also to the Standard Model fermions, and induce similar formulae for their masses?

\paragraph{Criticality:}
The first motivation is provided by the criticality of the Standard Model, whereby 
the top-quark mass may be related to the Higgs mass, as proposed by Froggatt and Nielsen\cite{Froggatt:1995rt}.
This, in turn, implies, via~\eqref{e:AbelDienes2}, that the mass of the top also could be related to
the cosmological constant --- because the Higgs mass is. 
Again by dimensional analysis, as in~\eqref{e:AbelDienes2}--\eqref{e:mH-BHM}, the analogous fermionic formulae are expected to be of the form $m_\psi \sim g_s M_s$,
up to the multiplicative coefficients implied by stringy modular invariance. 
This suggests a seesaw formula akin to the one for the Higgs mass~\eqref{e:mH-BHM},
however with appropriate UV and IR
scales.\footnote{This indeed follows Weinberg's general idea, ``in some leading approximation the only quarks and leptons with nonzero mass are those of the third generation, the tau, top, and bottom, with the other lepton and quark masses arising from some sort of radiative correction''\cite{Weinberg:2020zba}, except that the lower fermion masses are here generated by variants of the T-duality seesaw mechanism from a stringy non-perturbative effect; see \SS\,\ref{s:mStr}.} The claim here is that such seesaw formulae relate seemingly independent fermionic masses (in different generations) in the Standard Model.
In essence, this reasoning provides for the origin of different generations, starting from the heaviest fermions, and predicts that there can exist no heavier generations of Standard Model fermions.

However, what UV and IR scales are appropriate in such fermionic seesaw formulae?
To this end, recall that the masses of the charged fermions ($m_t$, $m_b$ and $m_\t$) are related via the RG equations for the heaviest fermions in explicitly computable stringy models\cite{Faraggi:1991be}, and thus are natural candidates for the UV scales.
As to an appropriate IR scale, we present below an entropy argument that leads
to a scale typically related to the  standard QCD scale, but is smaller by an order magnitude; we call this the Bjorken-Zeldovich scale, $M_{BZ}\simeq7$\,MeV.
We then find (as observed by Bjorken in a completely different context\cite{Bjorken:2013aa}) that this $M_{BZ}$ can, with the masses of the heaviest charged fermions as the UV scale, parametrize the masses of the remaining charged fermions.

This observation can be properly justified only by
a computation of the bound of the partition function
of the Standard Model in the modular polarization, which by the already explicit computation of the cosmological constant is given by the volume of phase space. 
Relating then the number of phase space cells, in modular regularization, to the Bekenstein-like bound
with the UV scale given by the masses of the heaviest quarks and the heaviest lepton then reproduces Bjorken's expressions\cite{Bjorken:2013aa,rBJ-MM}.

\paragraph{Seesaw Structure:}
The foregoing discussion, including the stringy result~\eqref{e:AbelDienes2}, involves two types of formulae:
The geometric mean: $m<(m'\,{\sim}\,\sqrt{mM})<M$, is here implied by the non-commutative, symplectic structure of Born geometry, $\omega_{ab}$ in~\eqref{etaH0}.
The ``seesaw-light,'' $m''\sim(m^2/M)<m<M$, is familiar from neutrino physics and is here of the T-duality type, implied by the bi-orthogonal structure of Born geometry, $\eta_{ab}$ in~\eqref{etaH0}. The presence of the double metric, $H_{ab}$ in~\eqref{etaH0}, is what allows the doubling of the heaviest mass in the first place.  This provides for three distinct masses and is, essentially, our key observation here.

This dovetails with the fact that there are three generations, and
meshes nicely with the present experimental constraints on
other generations of quarks and leptons. 
In what follows, the bounds on the charged fermion masses
take the form of these seesaw relations
(as used for the cosmological constant and also for the Higgs mass):
with $M_{UV}$ identified with the heaviest mass,
the lighter copies are $M_{IR}$-multiples of numerical factors that are solely the square-root of ratios of the UV and IR scales, or the other way around:
\begin{subequations}
 \label{e:twoRoots}
 \begin{alignat}9
    M_{IR} \sqrt{\frac{M_{UV}}{M_{IR}}}&=\sqrt{M_{IR}M_{UV}},
     &\qquad&\text{ for the middle, and} \label{e:twoRoots1}\\
    M_{IR} \sqrt{\frac{M_{IR}}{M_{UV}}}
     &=\sqrt{\Big(\frac{M_{IR}^2}{M_{UV}}\Big)\,M_{IR}},
     &\qquad& \text{for the lightest.}\label{e:twoRoots2}
 \end{alignat}
\end{subequations}
With the UV and IR scales as reasoned above, one expects the numerical factors in~\eqref{e:twoRoots} to be square-roots of their ratios. Analogously, the dominant $\cX$-term in~\eqref{e:AbelDienes2} gives $m_H\sim g_s^2M_P=\sqrt{M_\L/M_P}\,M_P=M_\L\sqrt{M_P/M_\L}$. Higher powers of these square-root factors then correspond to higher powers of $g_s^2$, and are expected as (string-perturbative) {\em corrections\/} to~\eqref{e:AbelDienes2}.\footnote{Also, the evident $g_s\to g_s^{-1}$ map between~\eqref{e:twoRoots1} and~\eqref{e:twoRoots2} would seem to indicate that S-duality must be involved in an underlying stringy derivation of such formulae.} By the same token, higher powers of the square-root factors in~\eqref{e:twoRoots} are expected as corrections of these formulae.
 For example, the {\em standard} seesaw-formula from the original, neutrino physics,
\begin{equation}
  \frac{M_{IR}^2}{M_{UV}} = M_{IR}\,\Big(\frac{M_{IR}}{M_{UV}}\Big)
  = M_{IR}\,\Big(\sqrt{\frac{M_{IR}}{M_{UV}}}\Big)^2,
 \label{e:noRoots}
\end{equation}
features the {\em\/square\/} of the numerical factor in~\eqref{e:twoRoots2}, and is expected to corresponds to an additive correction to~\eqref{e:twoRoots2}.

Also, assume that a fermionic version of the stringy result~\eqref{e:AbelDienes} can be derived, with a corresponding insertion vev, $\vev{\cX_\j}$, proportional to the gauge charges of the fermion $\j$ as indeed {\em is\/} the case for the Higgs field\cite{Abel:2021tyt}.
Then:
 ({\small\bf1})~for charged leptons, $\cX_\j\neq0$, the second term in a \eqref{e:AbelDienes}-like formula dominates, and formula~\eqref{e:twoRoots1} follows.
 ({\small\bf2})~For chargeless neutrinos, $\cX_\j=0$, only the first term in a \eqref{e:AbelDienes}-like formula remains, and~\eqref{e:noRoots} follows.

The remaining (T-duality type) seesaw formula~\eqref{e:twoRoots2} 
stems from the central property of the zero modes of the metastring
captured by the action of the metaparticle~\eqref{mp1}, and 
especially the constraint 
$p{\cdot}\tilde{p} = \mu$.
This is precisely the second, ``seesaw-light'' type relation, where we identify $\mu=M_{BZ}^2$ and the size of the dual momentum space with the relevant charged fermion mass.
Unlike the first seesaw formula~\eqref{e:twoRoots1}, which essentially follows from the phase-space-like structure and so is associated with the symplectic form, this second seesaw formula~\eqref{e:twoRoots2} is induced by the bi-orthogonal structure of Born geometry.

Ideally, one would need a precise fermionic analogue of the Abel-Dienes formula for the Higgs mass in string theory\cite{Abel:2021tyt}.
In the absence of such explicit formulae, we identify key seesaw features that connect our approach to  Bjorken's observations\cite{Bjorken:2013aa,rBJ-MM}, which we then also extend to the CKM matrix (like Bjorken), but also to neutrinos and the PMNS matrix (in ways different from Bjorken).
We find it intriguing that
the same logic used for the computation of the cosmological constant extends,
first to the Higgs mass,
and then also to the masses of all quarks and leptons.

While these seesaw features  do appear to be  cohesive and coherent,
a firm proof  would require the formulation of an explicit treatment of the Standard Model (SM) as a {\em\/modular QFT\/}: Every SM field $\phi$ is defined over both spacetime and the dual (momentum-like) spacetime,
$\phi(x, \tilde{x})$, with an intrinsic non-commutativity\cite{Freidel:2017xsi,Freidel:2018apz}, 
$[x, \tilde{x}] = i \ell_{nc}^2$, where $\ell_{nc}$ is in principle contextual, and not necessarily the string length or the Planck length.\footnote{Both of these scales, $\ls$ and $l_P$, turned up naturally in the discussion in Section~\ref{s:mStr}, but note that the effective, physically relevant 4-dimensional Planck scale may be removed, even exponentially much, from the underlying fundamental scale, e.g., in the large class of models discussed in\cite{rBHM1,rBHM7,Berglund:2020qcu,Berglund:2021xlm}.} By construction, such a formulation would have a natural solution of the vacuum energy problem,
and then, we conjecture, would also lead to the formulae for the fermionic masses presented below.
Such a modular SM would thereby imply relations between masses of different fermion generations that are invisible to the standard QFT form of the SM.
Such a modular QFT form of the SM can be also embedded in the metastring, which suggest a completely new (and complementary)
view on the origin of the SM in string theory, as compared to the traditional one based on Calabi-Yau compactifications in the point-field limit QFT\cite{Polchinski:1998rq}.
This should indicate that there are missing concepts (modular spacetime, modular polarization, Born geometry,
modular fields, metaparticles and metastrings) in the usual approach, and that the introduction of these missing
concepts to the canonical approach would yield the results discussed in this paper.

Unlike the very concrete foregoing statements about the vacuum energy problem and the problem of the Higgs mass, our present discussion of fermion masses is just a working conjecture at the moment.
We now turn to the implementation of this general set-up by
following our recent presentation\cite{Minic:2023oty}.

\subsection{Masses}
\label{s:fMasses}
\paragraph{The Bjorken-Zeldovich scale:}
As pointed out by Bjorken \cite{Bjorken:2013aa}, the observed masses of quarks and leptons could
be all parameterized in terms of a new, $O(10\,\text{MeV})$-scale. This Bjorken-Zeldovich scale 
is given by the size of the universe and the Planck scale, $l$ and $l_P$:
\be
l_{BZ}^3 \sim l\,l_P^2 \overset{\sss\eqref{e:lcc=llP}}{\sim} l_{cc}^2\,l_P,
\quad\text{i.e.},\quad
M_{BZ}^3 \sim M_\L^2\,M_P,
\label{e:lBZ}
\ee
determined by the same IR and UV scales as the cosmological constant.
Most importantly, given~\eqref{e:NllP}, $N \sim l^2/l_P^2$, the
Bjorken-Zeldovich scale $l_{BZ}^3 \sim l\,l_P^2 \sim l^3/N $ and
so  $N \sim l^3/l_{BZ}^3$,
--- precisely as expected from {\em\/extensive\/} non-gravitational entropy.
Therefore, given:
 ({\small\bf1})~our $N$,
 ({\small\bf2})~the Bekenstein bound for gravitational degrees of freedom,
 ({\small\bf3})~the fact that in metastring theory the
matter and spacetime degrees of freedom are ``two sides of the same coin,''
 ({\small\bf4})~the extensive nature of entropy for the matter degrees of freedom
\be
N \sim l^3/l_{BZ}^3 \sim l^2/l_P^2,
\ee
we are able to {\em deduce} the Bjorken-Zeldovich scale,
$l_{BZ}^3 \sim l\,l_P^2 $ (corresponding to roughly $10$\,MeV, or equivalently, $10^{-14}$\,m). In what follows
we are careful about the numerical
values of $l$ and $l_P$ and will use
the value $M_{BZ}\simeq 7$\,MeV
for the Bjorken-Zeldovich scale,
as used by Bjorken (who in turn seems to have been inspired
by the work of the Oxford group\cite{Chan:2015bvx}).\footnote{Bjorken has discussed this scale in a radically different context of the MacDowell–Mansouri approach to gravity, and in particular, the Friedmann-Robertson-Walker cosmology in that formulation\cite{Bjorken:2013aa}. Our derivation of the Bjorken-Zeldovich scale is,
as far as we are aware, completely new.}
Thus, all three scales,
the cosmological constant mass scale, $M_\L$, 
the Higgs mass scale, $m_H$, as well as 
the Bjorken-Zeldovich scale, $M_{BZ}$, are all
ultimately determined in terms of the Hubble ($M$) and Planck mass ($M_P$) scales.

\paragraph{Quarks:}
To proceed, we use the masses of the heaviest fermions (the top- and the bottom-quark, as well as the tau-lepton) 
as the natural short distance scales. (For a concrete computation of these masses in a string theory model, see \cite{Faraggi:1991be}.) These masses serve as analogs of the UV scale in our Higgs mass formula~\eqref{e:mH-BHM}, whereas the
Bjorken-Zeldovich scale, $M_{BZ}\approx7$\,MeV, acts as the natural IR scale.
Note that the top mass $m_t$ is essentially tied to the Higgs
scale,\footnote{To this end, we cite the well-known argument based on criticality of the Standard Model that relates the masses of the top quark and the Higgs boson
\cite{Froggatt:1995rt} (see also \cite{Donoghue:2005cf,Khoury:2022ish}, for landscape-motivated discussions).} which in turn is
given by the (geometric mean) seesaw formula of the vacuum energy scale and the Planck scale.
Thereby, the top quark mass is ultimately also given in terms of the Hubble and Planck mass scales.
Analogously to~\eqref{e:mH-BHM} for the Higgs mass, 
the (geometric mean) seesaw relation then produces the charm mass
in terms of $M_{BZ}$ and $m_t$ (cf.\ the observed value in parentheses\cite{rPDG22}):
\be
m_c \sim \sqrt{M_{BZ}\, m_t} = M_{BZ} \sqrt{\frac{m_t}{M_{BZ}}}
\sim 1.10~(1.27)\,\text{GeV}.
 \label{e:mch}
\ee
Next, using the bottom-quark mass scale\footnote{For example, explicit calculation in the stringy calculation\cite{Faraggi:1991be} ties, via RG equations, the mass of the top quark to the mass of the bottom quark and the tau lepton, and so are all ultimately determined by the Hubble and Planck mass scales.} 
(instead of $m_t$)
and the same Bjorken-Zeldovich scale as the characteristic vacuum energy scale of matter,
the same seesaw relation yields the mass of the strange quark
\be
m_s \sim \sqrt{M_{BZ}\, m_b} =
M_{BZ} \sqrt{\frac{m_b}{M_{BZ}}}
\sim 171~(93.4)\,\text{MeV}.
 \label{e:mst}
\ee
Bjorken estimates 
the up- and down-quark masses essentially at
the Bjorken-Zeldovich scale:
$m_u \sim M_{BZ}$ and  $m_d \sim M_{BZ}$,
but models the actual relation $m_d > m_u$ with {\em ad hoc\/} factors~\cite{Bjorken:2013aa}.
Independently, the masses of the lightest quarks may be
deduced from chiral perturbation theory as
$m_u \sim 2$\,MeV, $m_d \sim 5$\,MeV.
However, apart from non-commutativity that led to~\eqref{e:mch} and~\eqref{e:mst}, 
our seesaw structure reasoning above involves 
also the inherent metastring/metaparticle T-duality, 
which induces the familiar ``seesaw-light'' relation.
This then leads to the following estimates (actual values in parentheses\cite{rPDG22})
\be
m_u \sim M_{BZ}^2/m_c \sim
M_{BZ} \sqrt{\frac{M_{BZ}}{m_t}}
\sim 10^{-2}M_{BZ} \sim 10^{-1}~(2.16)\,\text{MeV}. \label{e:mup}
\ee
This estimate turns out too small (by a factor of about 50),
but is (importantly!) smaller than the down quark mass estimate (also too small by a factor of about 16), 
\be
m_d \sim M_{BZ}^2/m_s \sim
M_{BZ} \sqrt{\frac{M_{BZ}}{m_b}}
\sim 10^{-1} M_{BZ} \sim 1~(4.67)\,\text{MeV}.  \label{e:mdn}
\ee
The above reasoning thus automatically reproduces the 1st generation ``mass inversion'':
 $\eqref{e:mch}>\eqref{e:mst}$ but $\eqref{e:mup}<\eqref{e:mdn}$, which is necessary for the proton to be stable while the neutron decays.
Thus, given the heaviest, top and the bottom quark masses,
the two distinct seesaw type formulae (non-commutativity and T-duality) produce quite realistic estimates for the masses of the middle and the lightest quark generations.

\paragraph{Charged leptons:}
Turning to the charged leptons, the evident analogue of the top-quark is the tau-lepton.
From a naive stability analysis of the tau analogue of the hydrogen atom,
the mass of the tau is expected to be of the order of 
the mass of the nucleus, i.e. a GeV.
This is supported since the masses of the top, bottom quark and the tau lepton are all related by the RG equations, as in the calculation of\cite{Faraggi:1991be}.
With the tau mass as given (again, from the calculation of\cite{Faraggi:1991be}, and ultimately related to the Hubble and Planck mass scales, much as the top and bottom quark masses are), the (geometric mean) seesaw estimate of the muon mass is (actual value in parentheses\cite{rPDG22}):
\be
m_{\mu} \sim \sqrt{M_{BZ}\, m_{\tau}} =
M_{BZ} \sqrt{\frac{m_{\tau}}{M_{BZ}}}
\sim 112~(106)\,\text{MeV}.
\ee
Just as with quarks, the second (T-duality kind) seesaw relation then yields the electron mass, given the calculated muon mass
\be
m_e \sim \frac{M_{BZ}^2} {m_{\mu}} \sim
M_{BZ} \sqrt{\frac{M_{BZ}}{m_{\tau}}}
\sim 464~(511)\,\text{keV}.
\ee

This proposal thus reproduces 3 generations of charged Standard Model fermions and their masses,
by the framework of the dual space, the modular spacetime Born geometry, and ultimately the metastring, i.e., by the intrinsic non-commutativity and covariant T-duality of the metastring.
The masses of the two lighter generations are induced from the masses of 
the heaviest quarks and leptons, and 
are fixed by non-commutativity and T-duality,
in analogy with the reasoning that gives the Higgs mass and the cosmological constant.
All of these formulae are seesaw-like and contextual bounds.
All of them ultimately reduce to the
IR size of the universe and the UV Planck length.

\paragraph{Neutrinos:}
Turning to neutrino masses and following Weinberg's original dimension-5 operator proposal in the Standard Model\cite{Weinberg:1979sa} (implying Majorana masses as well),
we estimate the heaviest (``tau'') neutrino mass to be
\be
m_3 \sim m_H^2/M_{SM} \sim (10^{-1} - 10^{-2})\,\text{eV}, \label{e:m3}
\ee
where the SM scale $M_{SM}$ is given by a  ``would-be unification scale'' of
the SM couplings (as indicated by RG equations), $\sim10^{15-16}$\,GeV, 
and $m_H$ is the Higgs scale of around $1$\,TeV.
This heaviest mass would once again be ultimately given 
in terms of the Hubble and the Planck mass scales.
The middle (``muon'') neutrino mass is then given
by a (geometric mean) seesaw formula, involving a low vacuum energy scale.
Unlike all quarks and charged leptons, the neutrinos do not get
their masses from the Higgs mechanism, so the vacuum scale
cannot be $M_{BZ}$ (used for the charged fermions) and so must be 
the only other vacuum scale:
the cosmological vacuum scale associated with the cosmological constant~\eqref{e:lcc=llP}:
\be
m_2 \sim \sqrt{M_\L\, m_3} =
M_\L \sqrt{\frac{m_3}{M_\L}}
\sim (10^{-2} -10^{-2.5})\,\text{eV}. \label{e:m2}
\ee
By comparison, a similar mass value has been argued\cite{Aydemir:2017hyf}
to be natural by examining a dimension 6 analogue of Weinberg's operator, where a neutrino 
could acquires its mass from a fermionic condensate controlled by the
Bjorken-Zeldovich scale, with the electroweak cutoff scale:
$m_2 \sim M_{BZ}^3/m_{H}^2
      \underset{^{\smash{\eqref{e:mH-BHM}}}}{\overset{_{\eqref{e:lBZ}}}{\sim}} M_\L$.

Finally, the lightest (``electron'') neutrino mass is then
estimated by the (T-duality) seesaw formula
\be
m_1 \sim M_\L^2/m_2 \sim
M_\L \sqrt{\frac{M_\L}{m_{3}}}
\sim 10^{-4}\,\text{eV}. \label{e:m1}
\ee
According to the Particle Data Group\cite{rPDG22}, the sum of neutrino masses (coming from cosmology) is bounded by
$10^{-1}$\,eV, which is satisfied by the above normal hierarchy of neutrino masses. 
Also, these values satisfy the constraint on the square of the differences of masses, 
$10^{-2}\,\text{eV}^2 - 10^{-5}\,\text{eV}^2$, coming from neutrino oscillation experiments.

All these estimates for quark lepton and Higgs masses and for the cosmological constant mass scale are upper bounds; this bound for $m_u$ and $m_d$ essentially being given by $M_{BZ}$. 
We thus expect an attractor mechanism (as in \cite{Argyriadis:2019fwb}) 
that would ``glue'' all these 
values to their upper bounds. This would be consistent with the existence of a moduli-free self-dual fixed point in metastring theory \cite{Freidel:2015pka} that could explain
the apparent criticality of the Standard Model parameters \cite{Froggatt:1995rt}.
Finally, all these bounds on the fermion masses, much as the bounds on the cosmological constant and the Higgs mass, are determined in terms of the Hubble and the Planck mass scales.

\subsection{Fermion Mixing}
\label{s:fMixing}
Next we comment on the CKM and PMNS mixing matrices, generally given in the format\cite{rPDG22}
\begin{equation}
\begin{pmatrix}
 c_{12}\,c_{13} & s_{12}\,c_{13} & ~s_{13}\,e^{-i\d}~\\
 -s_{12}\,c_{23}-c_{12}\,s_{13}\,s_{23}e^{i\d}
        & -c_{12}\,c_{23}-s_{12}\,s_{13}\,s_{23}e^{i\d}
                 & c_{13}\,s_{23}\\
  s_{12}\,s_{23}-c_{12}\,s_{13}\,c_{23}\,e^{i\d}
        & -c_{12}\,s_{23}-s_{12}\,s_{13}\,c_{23}\,e^{i\d}
                 & c_{13}\,c_{23}\\
\end{pmatrix},
 \label{e:mixMat}
\end{equation}
where $c_{ij}\define\cos(\theta_{ij})$ and $s_{ij}\define\sin(\theta_{ij})$, with $0\leqslant\theta_{ij}\leqslant\pi/2$ and $\delta=\delta_{13}$. In particular:
\begin{equation}
    V_{CKM}=
\begin{pmatrix}
 V_{ud} & V_{us} & V_{ub}\\
 V_{cd} & V_{cs} & V_{cb}\\
 V_{td} & V_{ts} & V_{tb}\\
\end{pmatrix}
=
\begin{pmatrix}
 0.97373 & 0.2243 & 0.00382\\
 0.221 & 0.975 & 0.0408\\
 0.0086 & 0.0415 & 1.014\\
\end{pmatrix},
 \label{e:mixCKM}
\end{equation}
with experimental errors in the last digits\cite{rPDG22}.

\paragraph{The CKM Matrix:}
Quark mixing can be usefully parametrized (in close analogy with
Bjorken's parametrization \cite{Bjorken:2013aa}) by taking the following three crucial
entries (which are equivalent to the knowledge of
the above three independent angles) and by writing them in terms of the relevant vacuum scale for
the case of quark masses (the Bjorken-Zeldovich scale $M_{BZ}$) and the
seesaw like expression involving the masses of the bottom quark and 
the down, strange and bottom quarks, respectively.
The formulae listed below involve the product of two seesaw factors, that is, the product of two square root factors of the ratio of two scales,
as explained in the beginning of this section.
Explicitly, we have the following pattern, with Bjorken's values given in parentheses:
\begin{alignat}9
|V_{cb}| &\sim \frac{M_{BZ}}{\sqrt{m_b\,m_d}}
&&\sim \sqrt{\frac{M_{BZ}}{{m_b}}} \sqrt{\frac{M_{BZ}}{{m_d}}}
 &&\sim 0.050 \quad (0.041),\qquad
 &&(\leadsto\theta_{23})
\label{e:Vcb}
\intertext{(essentially, $(M_{BZ}/m_b)^{1/4}$)
as well as}
|V_{td}| &\sim \frac{M_{BZ}}{\sqrt{m_b\,m_s}}
&&\sim \sqrt{\frac{M_{BZ}}{{m_b}}} \sqrt{\frac{M_{BZ}}{{m_s}}}
 &&\sim 0.011 \quad (0.008)
 &&(\leadsto\theta_{12})
\label{e:Vtd}
\intertext{(essentially, $(M_{BZ}/m_b)^{3/4}$)
and finally}
|V_{ub}| &\sim \frac{M_{BZ}}{\sqrt{m_b\,m_b}}
&&\sim \sqrt{\frac{M_{BZ}}{{m_b}}} \sqrt{\frac{M_{BZ}}{{m_b}}}
 &&\sim 0.002 \quad (0.003)
 &&(\leadsto\theta_{13})
\label{e:Vub}
\end{alignat}
(essentially, $M_{BZ}/m_b$).
Comparing with~\eqref{e:mixMat} and~\eqref{e:mixCKM}: {\em first,} $\theta_{13}$ is determined from~\eqref{e:Vub}; with that, $\theta_{23}$ is determined from~\eqref{e:Vcb} {\em second,} and with those, $\theta_{12}$ is determined from~\eqref{e:Vtd}.
As in Bjorken's parametrization (where~\eqref{e:Vcb} is replaced by $(M_{BZ}/m_b)^{1/2}$), 
these values are quite good when
compared to experiment, except perhaps for the
first value which was dependent on the value of the down quark, that is, according to our prescription off by an approximate factor of 
$10$ from the observed value.

\paragraph{The PMNS Matrix:}
This is parametrized following exactly the
same pattern as the CKM matrix, with the following replacement:
As in the discussion of the neutrino masses, $M_{BZ}$ is replaced by $M_\L$. 
Otherwise, the
replacement of the bottom mass by $m_3$ and the strange mass
by $m_2$ and finally, the down mass by $m_1$ is obvious.
We also take into account that $m_3$ is known up to a factor of $1/10$
in the above formula for the heaviest neutrino mass (we include the observed data in parentheses\cite{Esteban:2020aa,rPDG22})
\be
|U_{\mu 3}| \sim \frac{M_\L}{\sqrt{m_3 m_1}} 
\sim \sqrt{\frac{M_{\Lambda}}{{m_3}}} \sqrt{\frac{M_{\Lambda}}{{m_1}}}
\sim 0.50, \quad (0.63)
\label{e:Um3}
\ee
(essentially, 
$(M_\L/m_3\overset{\sss\eqref{e:m3}}{\sim} M_{SM}/M_P)^{1/4}$) as well as
\be
|U_{\tau 1}| \sim \frac{M_\L}{\sqrt{m_3 m_2}} 
\sim \sqrt{\frac{M_{\Lambda}}{{m_3}}} \sqrt{\frac{M_{\Lambda}}{{m_2}}}
\sim 0.13, \quad (0.26) 
\label{e:Ut1}
\ee
(essentially, $(M_\L/m_3\sim  M_{SM}/M_P)^{3/4}$)
and finally
\be
|U_{e3}| \sim \frac{M_\L}{\sqrt{m_3 m_3}} 
\sim \sqrt{\frac{M_{\Lambda}}{{m_3}}} \sqrt{\frac{M_{\Lambda}}{{m_3}}}
\sim 0.06, \quad (0.14)
\label{e:Ue3}
\ee
(essentially, $M_\L/m_3\sim  M_{SM}/M_P$).
These values, are to first order, quite good when compared to
the observed data \cite{rPDG22}.
\footnote{Bjorken has different masses for neutrinos and his PMNS matrix is of the tri-bimaximal type \cite{rBJ-MM}. In his treatment of the neutrino sector the characteristic scale is still $M_{BZ}$.}

We observe that even though the numerical values of the CKM and PMNS matrices
are quite different, the underlying pattern in both cases is the same.
(However, we do not know the precise origin of this
underlying pattern. For that we would need a fermionic analog of the Abel-Dienes stringy formula for the Higgs mass\cite{Abel:2021tyt}.)
The crucial difference stems 
from the appearance of $M_\L$ for neutrinos in place of $M_{BZ}$ for quarks and charged leptons.
Also the heaviest neutrino is determined by the Weinberg dimension 5 operator,
but the other two masses follow the patterns found in the case of quarks and charged leptons,
except for the crucial $M_{BZ}\to M_\L$ replacement.
In our approach the CP violating phases would come from the SM calculation and the dual SM sector as
well from the intrinsic CP violation of quantum gravity in the non-perturbative metastring theory.

In conclusion to this section, note that the Standard Model of the observed kind 
(and not its SuSy extension) could be obtained by understanding the gauge groups as
general quantum phases. 
Recall that 
the $E_8$ prediction
of string theory as an overarching gauge group could be understood from the point of view of octo-octonionic geometry, which by dimensional reduction
to real-octonionic geometry
gives the geometry of the unique octonionic quantum theory captured by the octonionic projective geometry
with the isometry group of $F_4/SO(9)$, whereas $SO(9)$ is the general quantum phase,
that upon its compatibility with the 4 dimensional Poincare group leads to the Standard Model gauge group\cite{Gunaydin:1974fb,Gunaydin:1973rs}.
This  is different from the usual GUT logic, but it points to a possible robustness of the Standard Model group
(and its dual Standard Model of the dark sector, coming from the other $E_8$ in heterotic string theory).
Note that this fits within the metastring formulation, because the heterotic string is constructed from the bosonic string in 26 dimensions\cite{Casher:1985ra},
and the metastring is just its T-dually covariant chiral (phase space-like) formulation.

Finally, from this bottom-up point of view (our discussion has been in some
sense top-down) a modular quantization of the SM, coupled to the modular extension of general relativity, should give 
the structure that is implied by the top-down quantum gravity/string theory
approach. We hope to return to such a deeper investigation of this approach in the future.

\section{Conclusions and Outlook} 
\label{s:Coda}
In this paper we have presented in more detail the new argument regarding the calculation of the cosmological constant, $\L_{cc}$.
In particular, we have discussed the seesaw formula for the associated length scale, $l_{cc}$~\eqref{e:lcc=llP}, which exhibits UV/IR mixing, and that $l_{cc}$ is radiatively stable and natural.
We have also shown how the logic of this resolution of the cosmological constant problem, with input from the Abel-Dienes stringy calculation, extends naturally to the Higgs mass~\eqref{e:mH-BHM}.
Finally, we have shown that the same idea applies to the masses and mixing of quarks and leptons.
One important new ingredient in this reasoning is quantum contextuality (instead of the standard anthropic reasoning) which stems from the string/modular QFT vacuum being governed by Born geometry based on the modular phase space view of quantum spacetime \`a la\cite{Freidel:2022ryr}.
The interplay of phase space (and Born geometry), the Bekenstein bound, the mixing between ultraviolet (UV) and infrared (IR) physics and modular invariance in string theory (in its intrinsically non-commutative, metastring formulation) was emphasized
throughout the article. 

In particular, we have repeatedly stressed the purely stringy or quantum-gravity-related effects which are fundamentally rooted in the properties of quantum spacetime.
Such effects are not part of the usual EFT lore, largely because EFT lives in classical spacetime.
This might be disturbing to some readers, given the success of EFT. Consequently, we have argued that EFT results, dressed up with holography, can be recovered in a singular limit of our computation of the vacuum energy. Given the fact that the usual compactification approach to string theory, and the associated string landscape and swampland\cite{Polchinski:2006gy,Agmon:2022thq,Berglund:2022qsb}, are closely tied to EFT, we conjecture that the application of holography in that context, and a seesaw relation between what are usually considered UV and IR cut-offs in EFT\cite{Cohen:1998zx}, could lead to a top-down realization of our computations and results, at a critical self-dual point (without moduli) which would hide the fundamental aspects of our discussion: modular spacetime, Born geometry and the metastring formulation.
The four-dimensional nature of our discussion may in this approach be related to the fundamental properties of strings at high temperature in the early universe\cite{Brandenberger:1988aj}. This could be then generalized to the computations of the Higgs mass and the masses and mixing matrices of quarks and leptons,
as discussed in this paper, revealing, perhaps, an attractor mechanism in string landscape and swampland.
In conclusion,
we list some further phenomenological implications of our work.

Our calculation of the cosmological constant introduces a new quantum number~\eqref{e:N}, $N$, which may be probed in gravitational waves, via gravitational wave ``echoes'': In particular, our result~\eqref{e:NllP} relates the number of phase space boxes to the Bekenstein bound, $N \sim l^2/l_P^2$. It can therefore be used for black holes, where $l\to l_{bh}$ is the size of the black hole horizon, where it is naturally related\cite{Bekenstein:1995ju}. In this case, the relevant quantization number, $N_{bh} \sim l_{bh}^2/l_P^2$, for black holes is of the order of $10^{80}$, and a possible observable feature of this quantization, $l_{bh}^2 \sim N l_P^2$, might be via the ``gravitational wave echoes'' \cite{Wang:2019rcf, Cardoso:2019apo} --- in the ``quantum chaos'' phase, given the enormous value of $N$.

Notably, the crucial seesaw formula, $\delta \sim \sqrt{l_{IR}\, l_{UV}}$, 
(with a characteristic IR length-scale $l_{IR}$ and the characteristic UV length-scale $l_{UV}$) appears in other related contexts, such as the gravitational wave interferometry probes of quantum gravity; see, for example, \cite{Verlinde:2019xfb}. In that context, our vacuum energy calculation can be performed on the level of the causal diamond of the interferometer ($l_{IR}$ being given by the length of the interferometer and $l_{UV}$ by the Planck length), leading to the same seesaw formula, except interpreted as an empirical probe of modular spacetime. In that case, instead of the characteristic IR length scale of $10^{-4} m$, one would have the scale of $10^{-16} m$.
        
Furthermore, seesaw formulae for the SM fermion masses follow from the same reasoning that lead to the cosmological constant~\eqref{e:lcc=llP} and the Higgs mass~\eqref{e:mH-BHM} seesaw formulae. In that situation, a new Bjorken-Zeldovich scale can be deduced (by analogous reasoning) which enters into Bjorken-like seesaw formulae for all masses of charged elementary fermions.\footnote{In a fermionic \eqref{e:AbelDienes2}-like formula, the $\cX_\j$-insertion term must be proportional to gauge charges, and so is absent for neutrinos. In any explicit model-dependent calculation such as\cite{Faraggi:1991be}, the RG equations ``tie'' the heaviest charged fermion masses to the electroweak scale, while for neutrinos the relevant RG equations extend the UV scale to $M_{SM}\sim10^{15-16}$\,GeV.} This approach seems to proffer a new view on the observed three generations of quarks and leptons as well as their respective mixing matrices.
Here we point out an analogy with critical phenomena and the mean field/Landau-Ginzburg (LG) approach which gives ``square root type'' formulae, or the critical index of 1/2, without any anomalous dimensions, and which are, in turn, introduced by a more precise renormalization group (RG) treatment of the LG like description. In our case, the analogue of LG is the modular field theory extension of the SM and gravity. Our formulae should therefore be understood in the ``mean field theory sense''. In the context of modular field theory (a consistent limit of metastring theory) we also expect a {\em double\/} RG that is sensitive both to UV and IR scales\cite{Freidel:2017xsi}. Also, all these formulae can be rewritten ultimately using only the Hubble (IR) scale $M$ and the Planck (UV) scale $M_P$. These are the only two scales that appear in all expressions for the cosmological constant, the Higgs mass and the masses and mixing of quarks and leptons.

We also point out that in the visible sector we ultimately have to work with
modular fields $\phi(x, \tilde x)$\cite{Freidel:2017nhg}. This is not so
in the Standard Model (SM) as it is understood at the moment,
but is implied by the modular polarization
and our argument about the bounds of fermion masses.
Thus, the modular SM fields should know about
the symplectic and also the biorthogonal 
structures associated with $x$ and $\tilde x$.
(This suggests a kind of generalized mirror symmetry in the visible sector.)
This is what induces two distinct seesaw formulae
(one non-commutative/symplectic, and one T-dual/biorthogonal),
naturally yielding three generations in $x$-spacetime 
(a heavy fermion and its two seesaw copies).
The invisible (dark) sector is spanned by the dual 
fields $\tilde \phi (x, \tilde x)$, which
may well be subject to a third
quantization indeterminacy because of an induced 
non-commutativity between visible $\phi$ and 
dual (invisible/dark) $\tilde\phi$ fields.
Thereby, while one may be able to deduce
the bounds on the parameters of the Standard Model (SM) in
string theory/quantum gravity, the ensuing indeterminacy in the parameters 
of the dual Standard Model (the dark sector)
should then be reciprocal to the relatively high precision (small indeterminacy) 
of the SM parameters.
(This would be in the spirit of the old ``third quantization'' proposal\cite{Strominger:1988si}). 

We emphasize that metaparticles\cite{Freidel:2018apz}
(zero modes of the metastring) represent a generic prediction of metastring theory and the dark matter sector can be seen as coming from a dual Standard Model with a dynamics that is entangled/correlated with the visible Standard Model\cite{rBHM10}.
The dark matter degrees of freedom are thus tied to the dual particles to the visible SM particles\cite{Berglund:2022qsb}.
Furthermore, this approach shows dark energy (modeled as the cosmological constant) to be the curvature of the dual spacetime, and naturally small\cite{Berglund:2022qsb}.
The natural relation between the dark matter and dark energy sectors in our formulation, as well as the relation between the visible and dark sectors, offers, apart from quantum contextuality, a new view on the coincidence problem in cosmology\cite{Polchinski:2006gy}.

Finally, perhaps the most dramatic prediction of dynamical Born geometry implies ``gravitizing the quantum''\cite{Freidel:2014qna,Berglund:2022qcc,Berglund:2022skk}, and the presence of intrinsic and irreducible triple (and higher order) interference 
(\`a la Sorkin\cite{Sorkin:1994dt}) 
in the presence of gravity; see\cite{Berglund:2023vrm}. This would be a new quantum probe of quantum spacetime and a new avenue
in quantum gravity phenomenology\cite{Hossenfelder:2012jw,Addazi:2021xuf}.

In conclusion, responding to Weinberg's closing thoughts\cite{Weinberg:2020zba}, we believe that
the ideas, arguments, calculations and results discussed in this paper 
will
inspire the much needed more detailed work devoted to developing all these  fundamental aspects of particle physics, quantum gravity and quantum gravity phenomenology.

\paragraph{Acknowledgments:}
First and foremost, we are grateful to Prof. Steven Weinberg, our teacher, role model and great influence, certainly well beyond our shared time in his Theory Group at UT, Austin, as graduate students (PB and DM) and as a postdoc (TH). We should like to hope that he would be pleased with our efforts reported here.
Many thanks to L.~Freidel, J.~Kowalski-Glikman and R.\,G.~Leigh as well as A.~Geraci, and D.~Mattingly  
for insightful collaborations and illuminating discussions.
We also thank S.~Abel, N.~Afshordi, L.~Boyle, T.~Curtright, K.~Dienes, P.~Draper, E.~Guendelman, L.~Hardy, Y.~H.~He, J.~Heckman, T.~Jacobson,
T.~Kephart, J.~Khoury, L.~Lehner, E.~Livine, A.~Mazumdar, L.~McAllister, H.~P\"{a}s, P.~Ramond, M.~M.~Sheikh-Jabbari,
D.~Stojkovic, T.~Takeuchi and N.~Turok for interesting questions and comments. 
PB is supported in part by the Department of Energy grant DE-SC0020220, and
thanks the
Simons Center for Geometry and Physics and the CERN Theory Group, for their hospitality.
TH is grateful to the Department of Physics, University of Maryland, and the Physics
Department of the University of Novi Sad, Serbia, for recurring hospitality and resources.
DM is supported by the  Julian  Schwinger Foundation and
the U.S. Department of Energy (under contract DE-SC0020262) and he thanks Perimeter Institute for hospitality and support.
We also thank E.~Livine and the International Society for Quantum Gravity, T.~Curtright and E.~Guendelman and the BASIC conference series and S. Abel and the CERN meeting on Exotic approaches to the hierarchy problems for providing us with exciting opportunities to present this work.

\begingroup
\small\baselineskip=13pt \parskip=0pt plus2pt minus1pt
\bibliographystyle{utphys}
\addcontentsline{toc}{section}{\numberline{}References}
\bibliography{Weinberg}
\endgroup

\end{document}